\documentclass{PoS}

\usepackage{tablefootnote}

\newcommand\kms{km~s$^{-1}$}
\newcommand\msun{$M_\odot$}
\newcommand\mhi{$M_{\mathrm{HI}}$}

\def\be{\begin{equation}}
\def\ee{\end{equation}}
\def\a40{$\alpha$.40}

\def\arcsec{$^{\prime\prime}$}

\newcommand{\hi}{H\,{\sc i}}

\def\aj{l{AJ}}                   
\def\apj{{ApJ}}                 
\def\apjl{{ApJ}}                
\def\aap{{A\&A}}                
\def\mnras{{MNRAS}}             
\def\pasa{{PASA}}               
\def\nat{{Nature}}              

\title{MeerKAT and the Smallest Galaxies}

\ShortTitle{MeerKAT and the Smallest Galaxies}

\author{\speaker{Elizabeth A. K. Adams} and Tom A. Oosterloo\\
        ASTRON, the Netherlands Institute for Radio Astronomy\\
        E-mail: \email{adams@astron.nl}, \email{oosterloo@astron.nl}}

\abstract{On small scales, there exists a tension between observations of dwarf galaxies and predictions for low-mass dark matter halos from simulations (often referred to as the small scale crisis). This tension includes a mismatch in number count (e.g., the canonical "missing satellites problem") but also discrepancies in the internal structure (e.g., the "too big too fail" problem). 
More recently, observations have revealed that low-mass satellite galaxies appear to form structures
around their central galaxy ("planes of satellites") while these structures are not predicted
in cosmological simulations.
Detailed observations of low mass galaxies are critical for constraining the baryonic feedback processes that are used to alleviate these discrepancies. 
A particularly fruitful course is to study dwarf galaxies that have a substantial reservoir of neutral hydrogen (\hi).
These are the systems that are the most likely to be isolated,
helping to disentangle intrinsic properties from evolutionary effects, and \hi\ kinematics 
can offer an immediate constraint on the hosting dark matter halo.
Given MeerKAT's exquisite sensitivity, it can potentially contribute to these studies
of low-mass \hi-rich dwarf galaxies that will help resolve  the small scale crisis. 
The current large \hi\ surveys are not designed for these studies,
but will still manage to detect a sample of galaxies with \hi\ masses below $10^7$ \msun\
comparable to the number of systems currently in the literature,
and will resolve  $\sim$15 systems with masses below $10^{8.5}$ \msun,
a critical regime for addressing which dark matter halos host low-mass galaxies.
We propose a thousand hour survey 
 of the Centaurus region, encompassing the M\,83 and Cen\,A galaxy groups,
which can robustly address key questions in understanding low-mass galaxies.
The central question to be addressed is: "How many galaxies equivalent to Leo T are there in the 
Centaurus region?"
Leo T is the lowest mass, gas-rich galaxy currently known; our proposed survey is designed to be
able to detect an object of similar \hi\ mass and linewidth throughout the volume of the Centuaurus
region, which will provide a full census of how many objects like this there are in a typical galaxy group.
In addition, planes of satellites have recently been identified around Cen\,A and our survey will be able
to address how far out these planes extend: are they structures concentrated only around the central galaxy
or are they connected to large scale structure?
}

\FullConference{MeerKAT Science: On the Pathway to the SKA\\
                 25-27 May, 2016\\
                 Stellenbosch, South Africa}

\begin{document}

\section{The importance of the smallest galaxies}

Studies of dwarf galaxies, especially those with a significant neutral hydrogen
(\hi) component, offer unique opportunities to test cosmological
and galaxy formation models.
Living in low-mass dark matter (DM) halos, these systems are the most
susceptible to feedback processes and disruption of their baryonic component.
Yet they survive, and field dwarfs even retain their gas reservoir.
Leo T is a prime example:
this galaxy has an \hi\ mass of $4.2 \times 10^5$ \msun\
\cite{AO2016}
and a stellar mass of only $10^5$ \msun\ \cite{2012ApJ...748...88W}.
This galaxy has recent star formation and is on the periphery of the
Milky Way: {\it how has it retained its neutral gas reservoir?}


The low-mass regime is also where predictions from cosmological simulations and observations
of galaxies
have the strongest divergence. Much of this can be explained by baryonic processes,
but this physics is implemented at the sub-grid level 
and predictions for low mass galaxies are extremely
sensitive to the resolution of the survey, e.g., \cite{2015MNRAS.453.1305W}.
Observations of low mass galaxies can be used to address a few key questions to help
constrain and interpret results from simulations:

 {\it What is the lowest mass galaxy that can form?}
There is a dearth of observed low-mass galaxies compared to theoretical
predictions for the number of low-mass DM halos.
In the Local Group, this is the canonical "missing satellites problem".
This discrepancy can be explained by baryonic processes resulting in the loss of
gas from the lowest mass halos, resulting in DM halos that lack an observable
counterpart. 
An important constraint for the simulations is determining if
 there an abrupt cutoff in galaxy mass and/or DM halo mass at which
there are no longer observable galaxies.
As the resolutions of simulations increase, they generally
produce lower mass galaxies \cite{2015MNRAS.453.1305W}, and
at the same time, as observations have increased in sensitivity, new low-mass galaxies have been discovered \cite{2015ApJ...813..109D}.
Observational studies to date have concentrated on
satellite galaxies around more massive central galaxies where environmental
processes are important.
An important step forward is to expand this work to low-mass field galaxies
as isolated systems. As these galaxies are typically gas-rich,
\hi\ surveys are an excellent means of both identifying and studying these galaxies.

{\it Which DM halos do low mass galaxies live in?}
The explanation for the lack of low mass galaxies requires that not all DM halos
host an observable galaxy. An important test of the simulations that reproduce
the required number of galaxies is to observationally confirm that observed dwarf galaxies
live in the DM halos that we expect.
The naive expectation is that more massive DM halos host galaxies while lower mass ones do not.
However, observations appear to show that observed dwarf galaxies live in less massive DM
halos than expected (the "too big to fail" problem).
This was originally seen in the satellites of the MW \cite{2012MNRAS.422.1203B} but has
recently been recognized to occur for field galaxies also;
this is important because
many of the proposed explanations for this discrepancy rely on interaction with
the central galaxy \cite{2015A&A...574A.113P}.
The kinematics of (marginally) resolved low-mass galaxies can be used to address this observationally,
by using rotation curve modelling or simply a last measured velocity and radius to constrain the
largest DM halo that could host a galaxy  \cite{2015A&A...574A.113P}.
Identifying low mass galaxies through resolved \hi\ surveys is an efficient observational strategy
as the kinematics are provided immediately,
and the identified systems are likely field galaxies.

 {\it Are low mass galaxies arranged in vast structures around more massive central galaxies?}
Structures of satellite galaxies have been identified around the Milky Way
\cite{2014ApJ...790...74P},
Andromeda \cite{2013Natur.493...62I} and Cen\,A \cite{2015ApJ...802L..25T}. 
These structures were not predicted from simulations, and it is not clear
if they are expected to be common in $\Lambda$CDM.
\hi\ surveys can help address the origin of these structures:
are they also seen in gas-rich dwarf galaxies?
These systems are more isolated and further from central galaxies;
this would indicate that these structures may be connected
to large scale structure.

\section{Prospectus for current surveys}

\begin{figure}
\centering
\includegraphics[width=\linewidth,keepaspectratio]{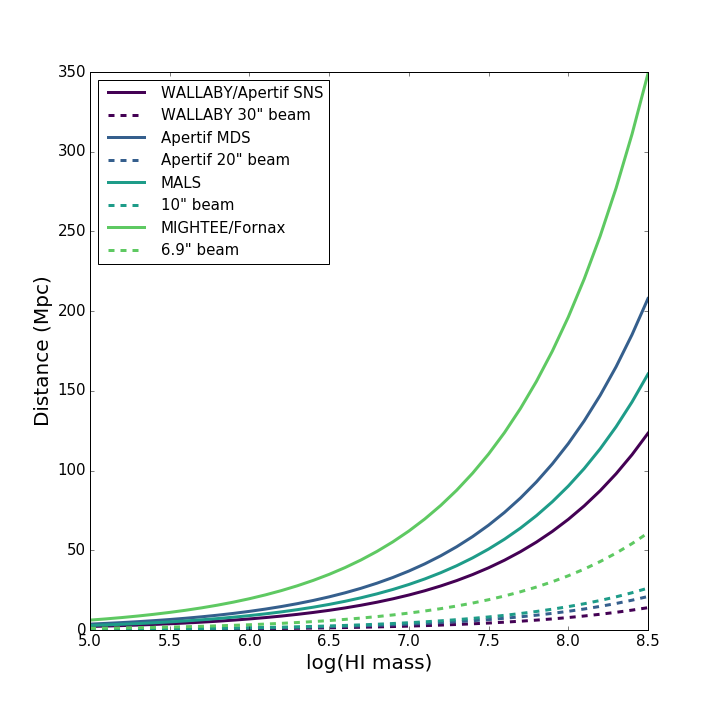}
\caption{Detection space for MeerKAT surveys and other surveys.}
\label{fig:detspace}
\end{figure}

Several of the large \hi\ imaging surveys already planned with MeerKAT have the potential
to detect and/or resolve galaxies in \hi\ mass ranges of key interest. 
The three surveys that we consider are MIGHTEE, Fornax and 
MALS\footnote{LADUMA has too small a footprint to probe a local volume where low mass galaxies
can be detected/resolved. MHONGOOSE observations do not reach out past the virial radius in most cases, where gas-rich galaxies are expected to be detected \cite{2014ApJ...795L...5S}.}.
These three surveys are described elsewhere in this volume; here we briefly summarize a few key parameters
for predicting their utility in detecting and resolving low mass galaxies.
For comparison, we also consider large \hi\ surveys planned with other SKA pathfinders;
these are WALLABY with the Australian SKA Pathfinder (ASKAP),
and the Shallow Northern Sky (SNS) and Medium Deep Survey (MDS) with Apertif,
a phased-array feed for the Westerbork Synthesis Radio Telescope.
Table \ref{tab:surveyparams} summarizes the relevant survey parameters.
Figure \ref{fig:detspace} shows the detection space for the MeerKAT surveys compared with the
other surveys. 
The sensitivity and resolution of MeerKAT allows
the planned surveys to detect and resolve low-mass \hi\ sources to further distances
than the surveys planned with other telescopes.

{\bf MIGHTEE}
is a simultaneous continuum and \hi\ survey of 20 deg$^2$.
MIGHTEE will have a sensitivity of 
$\sim$150 $\mu$Jy beam$^{-1}$ per 26 kHz channel when tapered to
 a 6.9\arcsec\ beam. For a linewidth of 20 \kms, this corresponds to a
 3-$\sigma$ sensitivity of $1.25 \times 10^{20}$ atoms cm$^{-2}$,
 or 1 \msun\ pc$^{-2}$.

{\bf The MeerKAT Fornax Survey}
is an \hi\ and continuum survey of the Fornax cluster.
This survey has a similar sensitivity to MIGHTEE with a footprint of
about half the size (12 deg$^2$).
An important key difference is that this field is specifically chosen
because of large scale structure: the Fornax cluster, located at a distance
of $\sim$20 Mpc.

{\bf MALS}
will be 1000 separate pointings
at bright radio sources to search for intervening \hi\ and OH absorption lines.
Simultaneously,
a shallow, sparsely sampled \hi\ imaging survey will be provided.
Since the survey consists of separate pointings, the sensitivity is not uniform.
We require a 3-$\sigma$ column density
limit of $1.25 \times 10^{20}$ atoms cm$^{-2}$
for a linewidth of 20 \kms. Then, for a 16\arcsec\ beam, the
effective area of MALS is 410 deg$^2$.
We approximate the point source sensitivity by using the average value between
the center of a pointing and the edge set by our column density sensitivity requirement.



{\bf WALLABY}
 is a shallow all-sky \hi\ survey with 
ASKAP. 
The stated goals of the survey are to survey 31,000 deg$^2$
at 30\arcsec\ resolution with a sensitivity of 1.6 mJy bm$^{-1}$
in a 4 \kms\ channel \cite{2012PASA...29..359K}.
For an intrinsic linewidth of 20 \kms, this corresponds to a 3-$\sigma$
column density limit of $5.3 \times 10^{19}$ atoms cm$^{-2}$.
In Table \ref{tab:surveyparams} we assume 50\% WALLABY sky coverage;
due to a likely decrease in sensitivity the mapping speed of 
ASKAP will decrease by a factor of $\sim$2,
and this represents that effect.
For a full WALLABY survey, 
the predictions can be multiplied by a factor of two.

{\bf The Apertif SNS}
 will
cover $\sim$3500 deg$^2$ 
to a similar sensitivity level as WALLABY.
The rms noise in a 20 \kms\ channel is expected to be 
0.65 mJy bm$^{-1}$ \cite{ASplan}.
The Apertif beam depends on the declination; for an average 20\arcsec\ beam
this corresponds to a 3-$\sigma$ column density limit of $1.1 \times 10^{20}$ atoms cm$^{-2}$.

{\bf The Apertif MDS}
is deeper observations
of a $\sim$450 deg$^2$ region, reaching 0.25 mJy bm$^{-1}$ for a 20 \kms\ channel,
or a 3-$\sigma$ column density limit of $4.1 \times 10^{19}$ atoms cm$^{-2}$. 
Part of the MDS footprint
includes the local overdensity of the Canes Venaciti loose groups,  CVn I and CVn II.

\begin{table}
\centering
\begin{tabular}{lllll}
\hline \hline
Survey & $S_{det}$ & $\theta$ & 3-$\sigma$ $N_{HI}$& FoV \\
	    & $\Delta \nu = 30$ \kms &  & $\Delta \nu = 20$ \kms & \\
            & mJy \kms  & \arcsec & atoms cm$^{-2}$ & deg$^2$\\
\hline
MIGHTEE & 11 & 6.9 & $1.25 \times 10^{20}$& 20 \\
Fornax & 11 & 6.9 & $1.25 \times 10^{20}$ & 12  \\
MALS & 52 & 16 & $1.25 \times 10^{20}$ & 410 \\
WALLABY & 88 & 30 & $5.3 \times 10^{19}$&  15000$^{a}$\\
Apertif SNS & 80 & 20 & $1.1 \times 10^{20}$& 3500 \\
Apertif MDS & 31  & 20 & $4.1 \times 10^{19}$&  450\\
\hline
Centaurus & 36 & 16 & $1.25 \times 10^{20}$ & 300\\
\hline
\end{tabular}
\caption{Summary of parameters for currently planned \hi\ surveys. 
Columns are: Survey, 5-$\sigma$ detection limit for a 30 \kms\ linewidth source
(3-$\sigma$ for a 84 \kms\ source), beam size
in arcseconds, 3-$\sigma$ column density limit for a linewidth of 20 \kms,
and survey footprint.
$^a$ This is 50\% of the nominal survey area; see text for a discussion of this choice.}
\label{tab:surveyparams}
\end{table}

We use the survey parameters to make predictions for the number of
sources that we expect the different surveys to both detect and resolve. 
We focus on two mass regimes of particular interest:
(a) $10^5 < M_{HI} < 10^7$ \msun\
as the lowest mass systems where we wish to determine
how many and which galaxies survive
and
(b) $10^7 < M_{HI} < 10^{8.5}$ \msun\ as the
mass range where more resolved \hi\ kinematics can help
address the too big to fail problem \cite{2015A&A...574A.113P}.
We integrate the \hi\ mass function (HIMF),
assuming the line flux detection limits in Table
\ref{tab:surveyparams}.
There we report the 5-$\sigma$ detection limit for a 30 \kms\ linewidth
unresolved source;
this is the typical width for a galaxy with \mhi $\sim 10^{6}$ \msun.
This same limit is the 3-$\sigma$ detection limit for a 84 \kms\ linewidth source,
a velocity width more typical for galaxies of $\sim 10^8$ \msun.
In general we use the HIMF from the ALFALFA \hi\ survey as a representative
field HIMF \cite{2010ApJ...723.1359M}. For the Apertif MDS we also use the
CVn HIMF \cite{2009MNRAS.400..743K} for the relevant part of the volume
to properly account for the known overdensity.
The CVn HIMF has a higher normalization than the ALFALFA HIMF as it
represents an overdense region of space. At the same time,
the slope of the low mass end is flatter, likely due to processing by the
environment and loss of \hi\ from some of the galaxies.
Fornax includes an overdensity, but, as a cluster,
is an even more extreme environment where
fragile low mass \hi\ objects are likely to be destroyed.
Determining the HIMF of Fornax is a science goal of that survey.
We
use the ALFALFA field HIMF as an approximation.
Because it does not account for the known overdensity,
more
low mass \hi\ sources are likely be detected than we predict.

In order to determine how many resolved objects are expected to be
detected, we again integrate the appropriate HIMF but set distance limits for each 
mass bin based on how far out an object can be resolved, rather than to where
it can be detected. In order to set these distance limits,
we use the \hi\ mass-diameter relation \cite{2016MNRAS.460.2143W}, 
 a tight correlation between the \hi\ mass of a galaxy and its
diameter at the 1 \msun\ pc$^{-2}$ ($1.25 \times 10^{20}$ atoms cm$^{-2}$) level.
For the roughest constraint of the resolved kinematics of galaxies,
we ask for 3 beams across a galaxy. This is not enough for detailed kinematic modelling
but may be enough to constrain a maximum velocity and the extent at which occurs,
allowing a constraint on the largest DM halo that may host a galaxy following the 
methodology of \cite{2015A&A...574A.113P}.
We also predict how many systems will be resolved by five or more beams;
these can be more fully modelled.

\begin{figure}
\centering
\includegraphics[width=\linewidth,keepaspectratio]{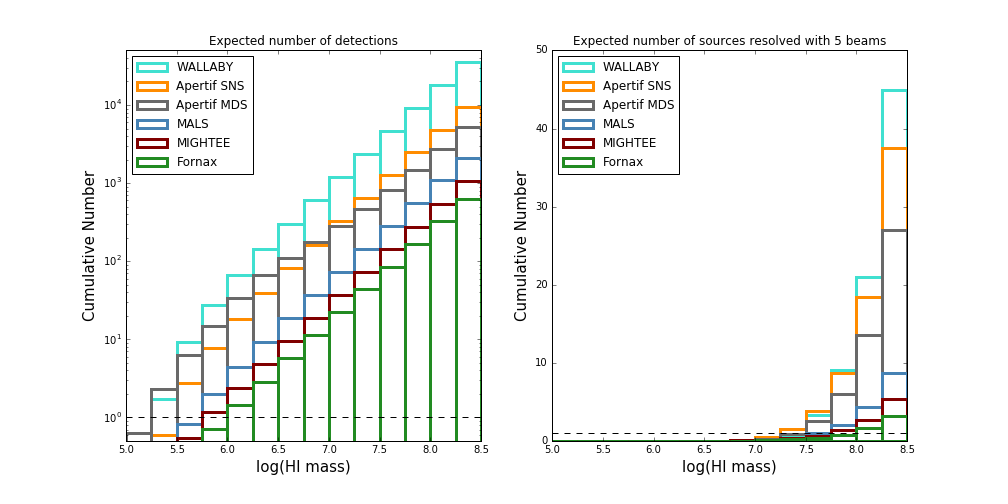}
\caption{Comparison of expected number of detections  and  resolved sources (>5 beams) 
at $D>3$ Mpc for the various
\hi\ surveys.}
\label{fig:surveypredictions}
\end{figure}

Table \ref{tab:surveypredictions} reports our predictions.
We note that these numbers should serve as a rough guide only.
We do not account for any scatter in the \hi\ mass-diameter relation
or for cosmic variance (except to discard systems
with $D<3$ Mpc;
objects at this distance or closer  may be at low velocities that are confused with foreground Milky Way
\hi\ emission).
This latter point point is especially important.
MIGHTEE and Fornax have small footprints and so the structure 
along the line of sight will heavily influence the
final number of detections.
The other surveys are relatively shallow and so can only detect and resolve low mass
galaxies at small distances, and structure in the local Universe will
strongly change their predictions.
Figure \ref{fig:surveypredictions} depicts the comparison in the number of expected
detected and resolved ($>$5 beams) sources for the various surveys.
While the MeerKAT surveys are more sensitive and have better resolution,
the shallow surveys planned with other telescopes are expected to detect
(and resolve) more objects due to their large footprint. 
The MeerKAT surveys may detect as many $\sim$60 sources with
\hi\ masses below 
$10^7$ \msun, approximately  double the number of sources
in this mass range in the literature. Given the rough value of these
estimates, the MeerKAT surveys should at least provide a sample of
the lowest mass \hi-selected dwarfs comparable to samples currently available.
The surveys will also well resolve $\sim$15 objects in the mass
range $10^7 - 10^{8.5}$ \msun,
an important parameter space
for determining which DM halos host observable galaxies.
A further $\sim$60 objects will be marginally resolved, which
may provide enough information to constrain their hosting DM halo.

\begin{table}
\centering
\begin{tabular}{llllllll}
\hline \hline
 &    & \multicolumn{3}{c}{$10^5 < M_{HI} < 10^{7}$ \msun} &
  \multicolumn{3}{c}{$10^{7} < M_{HI} < 10^{8.5}$ \msun}\\
Survey & HIMF & $N_{det}$ & $N_{3 beams}$ & $N_{5 beams}$
        & $N_{det}$ & $N_{3 beams}$ & $N_{5 beams}$\\
\hline
MIGHTEE & ALFALFA & 19 & 0.4 & 0.1 & 1000 &  25 & 5\\
Fornax & ALFALFA$^{a}$& 11 & 0.3 & 0.1 & 620 & 15 & 3\\
MALS & ALFALFA & 37 & 0.5  & 0 & 2100 & 41 & 9 \\
WALLABY & ALFALFA & 610 & 0.7 & 0  & 34000 & 220& 46 \\
SNS & ALFALFA & 156 & 2 & 0.1 & 9300 & 180  &  38\\
MDS & ALFALFA + CVn & 180  & 1.4 & 0.3  & 5000  &  70  & 27\\
\hline
Centaurus & CVn + ALFALFA & 63 (18) & 3 (3) & 0.5 (0.5) & 2600 (9) & 38 (9) & 13 (8)\\
\hline
\end{tabular}
\caption{Expectations for detections of low-mass galaxies for currently planned \hi\ surveys.
For the proposed Centaurus survey, numbers in parentheses
indicate the number of detections in only the Centaurus region, not including background sources.
$^{a}$Fornax is an overdense environment and the use of the ALFALFA HIMF is a conservative choice;
see text for a further discussion.}
\label{tab:surveypredictions}
\end{table}

\section{A dedicated MeerKAT project for the smallest galaxies}

MeerKAT has the sensitivity and resolution to greatly advance studies of the lowest
mass gas-rich galaxies.
The planned large surveys will produce a sample
of the lowest-mass \hi-rich dwarf galaxies and a sample of resolved
 dwarf galaxy kinematics
  comparable to currently existing samples.
However, this is not the prime focus of any of these surveys.
Here we propose a 1000 hour survey (850 hours of on-source time)
that would have a lasting legacy.

\subsection{Proposed survey}

We propose a survey of the Centaurus region, encompassing the M\,83 and Cen\,A
galaxy groups.
These are among the closest galaxy groups to us;
the Cen\,A group is at a distance of 3.76 Mpc, and the M83 group is
at a distance of 4.79 Mpc \cite{2007AJ....133..504K}.
The left panel of Figure \ref{fig:cengaldist} shows the distribution of distances
for galaxies in this structure.
Importantly for \hi\ studies, these groups have mean recessional velocities that
are clearly distinct from foreground Milky Way \hi\ emission;
the distribution of velocities of galaxies in these groups are shown
in the right panel of Figure \ref{fig:cengaldist}.
This both allows the detection of \hi\ sources without confusion with Galactic \hi\,
 and the assignment of a group
distance without any optical measurements.

\begin{figure}
\centering
\quad
\caption{Left: Histogram of distances of galaxies in Centaurus from \cite{2007AJ....133..504K}.
Right: Histogram of velocities of galaxies in Centaurus from \cite{2007AJ....133..504K}.
The systems are well separated from Galactic emission and systems with velocities of $\sim 600$ \kms\
can be directly associated with Centaurus without a distance measure.}
\label{fig:cengaldist}
\end{figure}

A core requirement for this survey is the ability to detect an object similar to Leo T over the volume
of the group. The \hi\ mass of Leo T is $4.2 \times 10^5$ \msun\
\cite{AO2016},
and the group edge is at 7 Mpc.
This corresponds to an \hi\ flux of 36.3 mJy \kms. To calculate
the time required to detect such a source, we take 30 \kms\ as a maximum linewidth
for such a low-mass object (the linewidth of Leo T is $\sim$20 \kms).
Using the updated specifications for MeerKAT ($SEFD =$ 424 Jy), and assuming a 1.5 loss of sensitivity
for a robust weighting scheme, we find that we require 1.7 hours of on-source integration time for a 5-$\sigma$ detection.


An important constraint is that we wish to be sensitive to the main
\hi\ disk of resolved systems.
We require that we be able to detect the 
column density level of $1.25 \times 10^{20}$ atoms cm$^{-2}$ (1 \msun\ pc$^{-2}$)
at the 3-$\sigma$ level for an intrinsic linewidth of 20 \kms.
For an on-source integration time of 1.7 hours, this implies a 16\arcsec\ beam or larger
be used for imaging.


We consider a survey footprint of 300 deg$^2$. This is a large
enough survey footprint to extend well beyond the virial radii of M\,83 and Cen\,A,
where we expect to detect few gas-rich galaxies
\cite{2014ApJ...795L...5S}, and to map the region between the two subgroups.
This footprint is shown in Figure \ref{fig:cengals}.
As discussed in the previous section, a large survey footprint is critical for increasing the number 
of low-mass sources detected.
In order to calculate the total time for the survey, 
we assume an effective field of view of 0.6 deg$^2$ for each pointing.
Then, it takes 850 hours of on-source integration time (or $\sim$1000
hours total including overhead) to map the Centaurus region
to the required sensitivity. The parameters of this proposed survey are included in Table \ref{tab:surveyparams}.

\begin{figure}
\centering
\caption{Galaxies in the Centaurus group from \cite{2007AJ....133..504K} (circles)
plus potential dwarf galaxy members (gray stars) from \cite{2015A&A...583A..79M,2016arXiv160504130M}.
The proposed 300 deg$^2$ footprint is shown by the red box.
}
\label{fig:cengals}
\end{figure}

\subsection{Expectations for the survey}
\begin{figure}
\centering
\includegraphics[width=\linewidth]{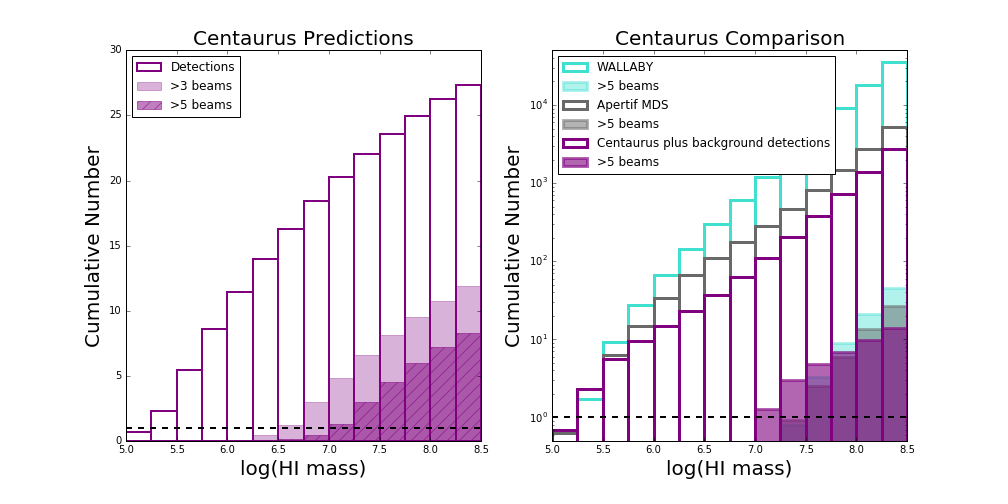}
\caption{Predictions for number of detections in the Centaurus  region.}
\label{fig:cenpredictions}
\end{figure}

Given our sensitivity limit, beam size, and footprint area,
we can predict how many sources we expect to detect and resolve.
We use the same mechanisms as in the previous section where we integrate
the HIMF using our survey parameters.
As this survey specifically targets an overdense region,
we use the CVn HIMF for the distance range of
the Centaurus group, 2-7 Mpc. 
We also make predictions for the detection of background sources at distances
greater than 7 Mpc using the ALFALFA HIMF.
Figure \ref{fig:cenpredictions} shows our predictions for the proposed survey.
In the left panel we show the predictions for the Centuaurs region only. 
In the right panel we compare the complete Centaurus survey (including background
sources) to other planned surveys.
We also report the number of sources we expect to detect in Table \ref{tab:surveypredictions}.

\subsection{Key science of the survey}
This proposed survey of the Centaurus region would detect and resolve a similar number
of galaxies as large-field surveys with other telescopes.
However, one key difference is that any detected \hi\ sources at velocities below 800 \kms\ can be associated
with Centaurus and thus have a group distance without optical follow-up. 
Good distances are key for studies of low mass galaxies \cite{2014ApJ...785....3M}.
Key science questions that will be addressed by this survey are:

{\it How many galaxies like Leo T exist in a typical galaxy group?}
By design, this survey will be able to detect an object with the \hi\ mass of
Leo T throughout the Centaurus region.
Leo T is on edge of detectability for SDSS, and its \hi\ content is almost confused in velocity-space
with the Galactic foreground.
Since the discovery of Leo T, further optical and \hi\ searches have only uncovered one similar
system, Leo P, which has an \hi\ mass of $8.1 \times 10^5$ \msun, a stellar mass of $5.6 \times 10^5$ \msun, and sits just outside the Local Group
at a distance of 1.62 Mpc
\cite{2015ApJ...812..158M}.
Given the observational difficulties in detecting these systems in our own galaxy group 
and the fact that Leo P was discovered as an \hi\ source
\cite{2013AJ....146...15G},
the best way to determine how common marginal systems like Leo T are may be 
by surveying an external galaxy group in \hi.

{\it How is the gas distributed in these systems and what is the potential for star formation?}
At a median distance of 4 Mpc, a 16\arcsec\ beam corresponds to a linear resolution of 300 pc.
This is comparable to the physical size of Leo T but would allow
more massive systems to be (slightly) resolved.
MeerKAT data has the large advantage that it can be imaged at multiple spatial scales.
Higher resolution maps can be produced, at the expense of column density sensitivity,
in order to localize high column density emission.
In addition, deeper follow-up observations with MeerKAT of the most
intriguing low-mass systems can be undertaken in order
to image the \hi\ at higher angular resolution.
Both Leo T and Leo P have a cool neutral medium component, evidence
for cooler dense gas and the potential for star formation \cite{2008MNRAS.384..535R,2014AJ....148...35B}.
It this common at this mass scale?

{\it Do the satellite structures extend to the field?} 
Recent observations show that the satellite galaxies around
Cen\,A are arranged in a plane \cite{2016arXiv160704024M}. With the addition of \hi\
observations, we can see if this structure extends to gas-rich galaxies that are
more removed from Cen\,A.
If this structure extends  further, that is a clue that it may be related to large scale structure.

{\it What halos do low mass galaxies live in?} 
We expect to well resolve  $\sim$10 sources in the Centaurus region below $10^{8.5}$ \msun.
These galaxies can serve as an important comparison sample for other \hi\ field surveys
as they live in a denser environment. Does
environment affect our predictions for the halos that host galaxies of a given mass
even when looking at an \hi\ rich population?

\section{Conclusions}

Studies of the smallest galaxies offer critical tests for cosmology and galaxy formation.
Currently planned MeerKAT surveys have the potential to 
double the number of low-mass dwarf galaxies with \mhi\ $< 10^7$ \msun.
They will also provide a number of (marginally) resolved systems below $10^{8.5}$ \msun\
comparable to current samples in the literature. The
resolved \hi\ data
allow a constraint on halos that galaxies live in which can be compared to results from abundance matching.
 This is a significant increase in the number of known and resolved systems in these mass ranges, but not competitive with
  \hi\ surveys planned for other SKA pathfinders.
We propose a survey of the Centaurus region, including the M\,83 and Cen\,A galaxy groups, which would
robustly address questions on low-mass galaxies, notably: "How common are systems like Leo T?".
This survey would also play an important role in the recent, open question of satellite galaxies forming vast structures;
the \hi\ observations would reveal if dwarf galaxies continue to from this structure
further from the central galaxy, implying that it may be connected to large scale structure.

\acknowledgments{We thank Kristine Spekkens for helpful discussion on the sensitivity limits of the various \hi\ survey and Paolo Serra for plots showing the effect of various weighting and tapering schemes on the sensitivity and restoring beam shape of MeerKAT.}




\providecommand{\href}[2]{#2}\begingroup\raggedright\endgroup

\end{document}